\documentclass[9pt,twocolumn,twoside]{osajnl}

\journal{ol} 

\setboolean{shortarticle}{true}

\title{Thermal control of Kerr microresonator soliton comb via an optical sideband}

\author[1]{Kenji Nishimoto}
\author[2,3]{Kaoru Minoshima}
\author[3]{Takeshi Yasui}
\author[3,4,*]{Naoya Kuse}

\affil[1]{Graduate School of Technology, Industrial and Social Sciences, Tokushima University, 2-1, Minami-Josanjima, Tokushima, Tokushima 770-8506, Japan}
\affil[2]{Graduate School of Informatics and Engineering, The University of Electro-Communications, 1-5-1 Chofugaoka, Chofu, Tokyo 182-8585, Japan}
\affil[3]{Institute of Post-LED Photonics, Tokushima University, 2-1, Minami-Josanjima, Tokushima, Tokushima 770-8506, Japan}
\affil[4]{PRESTO, Japan Science and Technology Agency, 4-1-8 Honcho, Kawaguchi, Saitama, 332-0012, Japan}
\affil[*]{Corresponding author: kuse.naoya@tokushima-u.ac.jp}

\begin{abstract}
We report the thermal control of a dissipative Kerr microresonator soliton comb via an optical sideband generated from an electro-optic modulator. Same as the previous reports using an independent auxiliary laser, our sideband-based (S-B) auxiliary light also enables to access a stable soliton comb and to reduce the phase noise of the soliton comb, greatly simplifying the setup with an auxiliary laser. More importantly, because of the intrinsically high frequency/phase correlation between the pump and S-B auxiliary light, the detuning between the pump and resonance frequency is automatically almost fixed, allowing the 18 times larger “effective” soliton existence range than the conventional method using an independent auxiliary laser, as well as the scanning of the soliton comb of more than 10 GHz without using microheaters. 
\end{abstract}

\setboolean{displaycopyright}{true}

\begin{document}

\maketitle

Microcombs are generated by seeding a continuous wave (CW) laser into a high-Q microresonator \cite{kippenberg2018dissipative}. Due to the CMOS compatible fabrication process, microcombs can be chip-scale and high-volume \cite{xiang2021laser, liu2021high}. A distinguished state of microcombs is the dissipative Kerr-microresonator soliton comb, hereafter called soliton comb, which is mode-locked and provides highly coherent ultrashort pulses, and low noise \cite{Herr_soliton}. Adding to applications similar to fiber combs such as optical atomic clocks and distance measurement, a new class of applications such as mm/THz waves generation \cite{zhang2019terahertz, tetsumoto2021optically} and photonic convolutional neural network \cite{feldmann2021parallel, xu202111} has been demonstrated by the soliton combs.

Different from conventional mode-locked lasers such as fiber-based combs, soliton combs are significantly affected by the thermal effect in microresonators because of high intracavity power and small mode-volume. The thermal effect emerges mainly in three cases. The first is the generation of soliton combs. When a soliton comb is generated, the sudden decrease of the intracavity power due to the transition from a chaotic to soliton comb causes the increase of the resonance frequency of a microresonator, which makes it difficult to keep the soliton comb \cite{chang2020ultra, moille2020dissipative}. The second is the phase noise of soliton combs. The thermal exchange between a microresonator and surrounding causes the thermal fluctuation, which couples to the fluctuation of the refractive index of the microresonator, affecting the phase noise of the repetition frequency of the soliton comb \cite{drake2020thermal, huang2019thermorefractive, nishimoto2020investigation}. The third is the control of the resonance frequency. The free spectral range (FSR) of a microresonator is shrunk via a microheater deposited on the microresonator through thermorefractive effect \cite{xue2016thermal, Gaeta_heater}, which is applied to the scanning of soliton combs \cite{kuse2020continuous, kuse2021frequency}. The first two are unfavorable and can be mitigated by using an auxiliary laser as follows \cite{zhang2019sub, lu2019deterministic, zhou2019soliton, wildi2019thermally, drake2020thermal}. The frequency of the auxiliary laser is set higher than that of the pump CW laser. While a soliton comb is generated, the pump CW laser and auxiliary laser are located at the red and blue side of the resonance frequency, respectively. Since the auxiliary is put closer to the resonance frequency, the thermal effect is mainly induced by the auxiliary laser. In this case, the auxiliary laser mitigates the thermal effect by counteracting the resonance frequency shift through the intracavity power change of the auxiliary laser. The auxiliary laser can be virtual, in which a pump CW laser also works as an auxiliary light by interacting with another resonance \cite{lei2021self}. On the contrary, the thermal effect plays an important role when soliton combs are scanned. The FSR of a microresonator can be changed by more than 0.1 \% via a microheater through the thermorefractive effect. To scan a soliton comb, both the frequency of the pump CW laser and resonance frequency of the microresonator are scanned simultaneously at the same amount not to change the detuning between the pump CW laser and resonance \cite{kuse2020continuous, kuse2021frequency}. 

Previously, an optical sideband generated from an electro-optic modulator (EOM) has been used to access a stable soliton comb with the extension of the "effective" soliton existence range \cite{wildi2019thermally}, which simplifies the reported setups \cite{zhang2019sub, lu2019deterministic, zhou2019soliton} with an independent CW laser as an auxiliary light and allows a robust operation. In this Letter, we unveil more advantages of the use of an optical sideband, showing the control of the above three thermal effects. In addition to the generation of a soliton comb with the careful investigation of the optimal properties (power and frequency) of the sideband, we utilize the large "effective" soliton existence for the frequency scanning of the soliton comb. Besides, the phase noise reduction of the soliton comb by the optical sideband is observed for the first time. In the experiments, we obtain the 100 times success of the generation of a single soliton comb in 100 tries and more than 20 dB phase noise reduction of the soliton comb. Moreover, we show the 18 times extension of the “effective” soliton existence range, which is utilized to scan the soliton comb, showing more than 10 GHz frequency scanning of the soliton comb without microheaters. 

Figure 1 shows the setup for the thermal control of the microresonator via an optical sideband (detail is shown in Fig. S1(a) in the supplementary material). An external-cavity diode laser (ECDL) as a master CW laser oscillating at around 1550 nm is split into two by a 50:50 optical coupler, which are used as a pump (red in Fig. 1) and an auxiliary (blue in Fig. 1) light. An optical sideband is generated from the auxiliary light by passing through a dual-parallel Mach-Zehnder modulator (DP-MZM), which is operated in the carrier-suppressed single-sideband (CS-SSB) mode. The frequency of the sideband-based (S-B) auxiliary light is separated from that of pump light by $f_m$ ($\sim$ 0 - 2 GHz) (Fig. 1). The amplified pump and S-B auxiliary light are coupled into the microresonator through lensed fibers. To separate the pump light from the S-B auxiliary light (or vice versa), an optical circulator is used before coupling into the microresonator. Although the pump and S-B auxiliary are coupled from the opposite direction, there is backscattering with a strength of about -15 dB. The loaded Q-factor and FSR are about $4.8 \times 10^5$ (see Fig. S1(d) in the supplementary material) and 1 THz, respectively. The generation of a soliton comb is confirmed by measuring the optical spectrum by an optical spectrum analyzer (OSA). The transient dynamics to generate the soliton comb is observed by monitoring the comb powers both from the pump and S-B auxiliary light, which requires an optical notch filter to reject the strong residual pump. The phase noise of the soliton comb is measured via a two-wavelength delayed self-heterodyne interferometer (TWDI) (detail is shown in Fig. S1(b) in the supplementary material) \cite{kuse2017electro, kuse2018photonic}.

\begin{figure}[!ht]
\centering
\fbox{\includegraphics[width=0.9\linewidth]{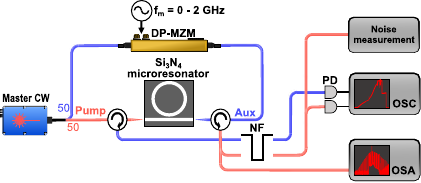}}
\caption{Schematic of the experimental setup. DP-MZM: dual-parallel Mach-Zehnder modulator, NF: notch filter, PD: photodetector, OSC: oscilloscope, OSA: optical spectrum analyzer.}
\end{figure}

In the experiments, $f_{\rm m}$ is set at 1100 MHz, and the pump and S-B auxiliary optical powers before the coupling ($P_{\rm pump}$ and $P_{\rm aux}$) are 150 mW and 100 mW, respectively, if not mentioned in the following. Note that a modulation-instability and chaotic combs are shown up at $P_{\rm aux}$ of 30 and 100 mW, respectively. An optimum pair of $f_{\rm m}$, $P_{\rm pump}$, and $P_{\rm aux}$ is found as shown in the supplementary material. Figure 2(a) shows the comb power from the pump (red) and S-B auxiliary (blue) light when the frequency of the master CW laser ($\nu_{\rm master}$) is swept from the red to the blue side of the resonance with the scan range and rate of 40 GHz and 10 Hz, respectively. Firstly, the comb power from the pump light increases, generating a chaotic comb. When the chaotic comb from the pump light transitions to a single soliton state at around 25 ms, the comb power from the pump (S-B auxiliary) light decreases (increases). Because of the opposite behavior between the comb power from the pump and S-B auxiliary light, the thermal resonance frequency shift is reduced, allowing the access of a stable single soliton comb even with the slow scan rate. When $\nu_{\rm master}$ is repetitively scanned by 100 times, the soliton comb was successfully generated for all the scanning, highlighted by the green band in Fig. 2(b). As shown in Fig. 2(c), the soliton comb has the comb mode spacing equal to one FSR of the microresonator and a smooth $sech^2$ envelope.

\begin{figure}[!ht]
\centering
\fbox{\includegraphics[width=0.9\linewidth]{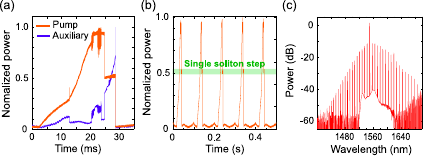}}
\caption{(a) Transient dynamics of the comb power from the pump (red) and S-B (blue) auxiliary light. (b) Comb power from the pump light when $\nu_{\rm master}$ is repetitively scanned. (c) Optical spectrum of the soliton comb.}
\end{figure}

One notable advantage of the auxiliary light from the optical sideband, instead of using another independent CW laser, is the significant extension of the “effective” soliton existence range. Here, we define the “effective” soliton existence range as to how large $\nu_{\rm master}$ can be swept without losing the soliton comb. As shown in Fig. 3(a), when the $P_{\rm aux}$ is 100 mW, the effective soliton existing range is about 18 GHz, which is 18 times larger than the case when an independent CW laser is used. The extension of the soliton existing range is due to the automatic and simultaneous frequency shift of the pump and S-B auxiliary light via the frequency shift of the master CW laser (Fig. 3(b)). When $\nu_{\rm master}$ moves towards red (blue), the intracavity power from the S-B auxiliary light becomes larger (smaller), inducing the red (blue) shift of the resonance frequency. As such, the detuning gets little change even when $\nu_{\rm master}$ is swept. The pushing force on the resonance by the S-B auxiliary light with $P_{\rm aux}$  = 10 mW is still effective, showing the effective soliton existence range of about 4 GHz. The detuning is indirectly inferred by measuring the optical spectra of the soliton comb. The detuning is proportional to the comb mode power \cite{yi2015soliton}. Figure 3(c) shows the optical spectra of the soliton combs at the blue and red edges of the effective soliton existing range. The optical spectra of the soliton comb with $P_{\rm aux}$ of 100 mW is very similar to those with $P_{\rm aux}$ of 0 mW, which indicates the detunings for $P_{\rm aux}$ of 0 and 100 mW are the almost same due to the automatic tracking of the thermal resonance frequency shift. As an indicator for the detuning change, we calculate $\cfrac{\delta P_{\rm comb}} {\delta _{\rm soliton}}$, where $\delta P_{\rm comb} \left(= \cfrac{P_{\rm red} - P_{\rm blue}}{P_{\rm blue}}\right)$ and $\delta _{\rm soliton}$ are the power change of the comb mode and amount of the frequency scan, respectively ($P_{\rm red (blue)}$: comb mode power indicaded by the arrows in Fig. 3(c) at the red (blue) edge). $\cfrac{\delta P_{\rm comb}} {\delta _{\rm soliton}}$ when $P_ {\rm aux}$ = 0 mW and 100 mW at the arrows in Fig. 3(c) are 1.01 GHz$^{-1}$ and 0.058 GHz$^{-1}$, respectively, which means the detuning change for $P_{\rm aux}$ = 100 mW is much slower than that for $P_{\rm aux}$ = 0 mW. $\nu_{\rm master}$ to keep the soliton comb is investigated by taking an optical beat with a stable single-frequency laser. Figure 3(d) shows the effective soliton existence range, depending on $P_{\rm aux}$. As $P_{\rm aux}$ decreases, the effective soliton existence range becomes narrower and shifts toward the high frequency. This is because of the less pushing force on the resonance. According to Fig. 3(d), the S-B auxiliary light can be turned off, if needed, by simultaneously reducing $P_ {\rm aux}$ and increasing $\nu_{\rm master}$ (e.g. by following the arrow in Fig. 3(d)). Turning off the S-B auxiliary light can avoid the detrimental effects (e.g. for the phase noise) from the reflection of the chaotic comb from the S-B auxiliary light and the S-B auxiliary light itself. The reduction of the optical power of the S-B auxiliary light also minimizes the electric power consumption of the system, which is important for power-efficient, chip-scale optical frequency comb. Alternatively, instead of changing $\nu_{\rm master}$, the power of the pump light can be increased to counteract the decrease of the intracavity power from the S-B auxiliary light. The detail of the alternative method is described in the supplemental material (Fig. S3).

\begin{figure}[!ht]
\centering
\fbox{\includegraphics[width=0.9\linewidth]{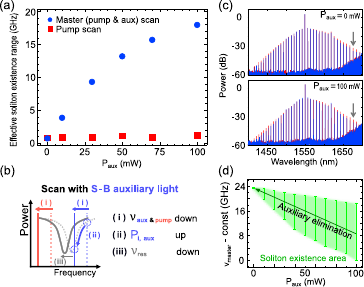}}
\caption{(a) Effective soliton existence range of the method with the S-B auxiliary light (blue) and independent auxiliary laser (red). (b) Illustration of the automatically fixed detuning between the pump light and resonance frequency. (c) Optical spectra of the soliton combs at the blue edge (blue) and red edge (red) of the soliton existence range when the $P_{\rm aux}$ is 0 mW (top) and 100 mW (bottom), respectively. The arrows show the comb modes used to calculate $\cfrac{\delta P_{\rm comb}} {\delta _{\rm soliton}}$. (d) Effective soliton existence area (green area), depending on $P_{\rm aux}$.}
\end{figure}

The automatic and simultaneous tracking of the thermal resonance frequency shift is also applied for the scanning of the soliton comb. Conventionally, the scanning of the soliton comb requires a microheater deposited on a microresonator \cite{xue2016thermal, kuse2020continuous, kuse2021frequency}. In addition, a feedback loop has to be employed to keep the detuning fixed while the soliton comb is scanned \cite{Vahala_power_mod16, Papp_PRL18}. Our thermal control method simplifies the system for the scanning of the soliton comb. In this experiment, a distributed feedback (DFB) laser is used as a master CW laser instead of the ECDL because the DFB laser allows a fast frequency scanning \cite{kuse2021frequency}. $P_{\rm aux}$ before the scanning is set at 100 mW. The scanning results are shown in Fig. 4. The soliton comb is scanned by more than 10 GHz up to the frequency scan rate of about 2 kHz. Above the frequency scan rate, the scan range is gradually decreased down to about 1.5 GHz for the scan rate of 200 kHz. To investigate the origin of the decrease of the scan range, the photothermal response of the optical microresonator was measured (the red curve in Fig. 4) by a method using a probe laser \cite{wildi2019thermally, liu2021high}. The detail is shown in the supplementary material. The photothermal response shows a similar trend to the decrease of the soliton scan range, although a deviation is observed at the scan rate from a few kHz to 50 kHz. The deviation is likely due to the power change caused by the current modulation of the DFB laser. The results suggest that the scan range/rate of the soliton comb via the thermal control method is limited by the photothermal response of our used Si$_3$N$_4$ microresonator. To achieve a fast scan rate without the decrease of the scan range, materials such as GaP for a fast photothermal response can be employed \cite{wilson2020integrated}. 

\begin{figure}[!ht]
\centering
\fbox{\includegraphics[width=0.8\linewidth]{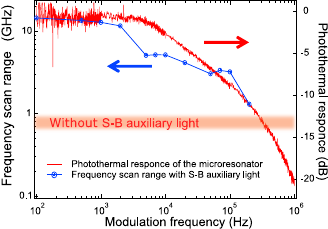}}
\caption{(Blue) The scan range of the soliton comb at different scan rates. (Red) Photothermal response of the microresonator.}
\end{figure}

Lastly, the S-B auxiliary light is used to reduce the phase noise of the repetition frequency of the soliton comb ($L_{\rm rep}$). $L_{\rm rep}$ is influenced by the thermorefractive noise of the microresonator due to the small mode volume. As explained in \cite{sun2017squeezing, drake2020thermal}, when the resonance shifts towards red (blue), the decrease (increase) of the intracavity power from the S-B auxiliary light in the microresonator causes the blue (red) shift of the resonance frequency. In other words, the S-B auxiliary light located at the blue side of the resonance passively counteracts the thermal random motion of the resonance frequency. In the experiment, $P_{\rm aux}$ is set as a sufficiently small value (< 30 mW) not to cause the detrimental effect from the chaotic comb of the S-B auxiliary light (see supplemental Fig. S8). Figure 5(a) shows $L_{\rm rep}$, measured by using the TWDI as shown in Fig. S1(b). Because of the recoil of the resonance frequency by the S-B auxiliary light, $L_{\rm rep}$ is suppressed by more than 20 dB in the wide range of the frequency offset up to 30 kHz. Above 30 kHz, the effect of the phase noise reduction is gradually reduced, although the phase noise reduction of 5 dB is still observed at the 1 MHz frequency offset. The gradual decrease of the effect would be related to the photothermal response of the microresonator as shown in Fig. 4(d). In addition to the phase noise reduction, our method with the S-B auxiliary light has another advantage; insensitivity to the frequency drift of the master CW laser because of the automatic tracking of the resonance frequency. $L_{\rm rep}$ is measured while changing $\nu_{\rm master}$. In Fig. 5(b), $L_{\rm rep}$ at the 10 kHz frequency offset is plotted. The optimum relative phase noise (< -65 dBc/Hz) is obtained in the broad frequency range ($\sim$ 7 GHz), which is about 17.5 times larger than the FWHM linewidth of the microresonator. As a comparison, the same experiment but with an independent auxiliary laser is also implemented as shown in Fig. S7, in which only the frequency of the auxiliary laser is changed. The frequency range for the optimum $L_{\rm rep}$ (< -65 dBc/Hz) is about 800 MHz, which is about 8.8 times smaller than the case with the S-B auxiliary light. According to these results, the advantage of using the S-B auxiliary light is not only to simplify the setup but also to be less sensitive to the frequency drift of the master CW laser. Also, the use of the S-B auxiliary laser might have another advantage due to the small relative phase noise between the pump and auxiliary light. However, we see little difference in the effect of phase noise reduction depending on the relative phase noise between the pump and auxiliary light (see supplemental Fig. S6), although the relative phase noise may be important when a high Q microresonator is used. 

\begin{figure}[!ht]
\centering
\fbox{\includegraphics[width=0.9\linewidth]{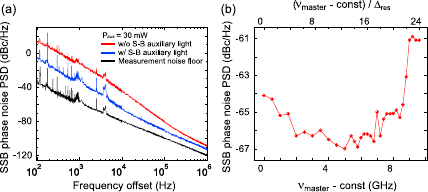}}
\caption{(a) $L_{\rm rep}$ with (blue) and without (red) the S-B auxiliary light. The measurement noise floor (black) is also shown. (b) $L_{\rm rep}$ at the 10 kHz frequency offset for the different $\nu_{\rm master}$. $\Delta _{\rm res}$ is the linewidth of the resonance frequency. 
}
\end{figure}

In conclusion, we demonstrated the thermal control of a soliton comb, in which an optical sideband generated from an EOM is used as an auxiliary laser. The demonstrated method not only simplified the setup with an independent auxiliary CW laser for the generation and $L_{\rm rep}$ of soliton comb, but also, more importantly, enabled the much larger “effective” soliton existence range due to the intrinsic phase correlation between the pump and S-B auxiliary light. The large “effective” soliton existence range would enable the use of soliton combs in harsh environments, where the frequency of the master CW laser and resonance frequency are significantly drifted. Also, because of the automatic and simultaneous tracking of the resonance frequency, the soliton comb was scanned by more than 10 GHz without microheaters, which is readily applied to massively parallel frequency-modulated CW (FMCW) LiDAR \cite{riemensberger2020massively}.

\medskip
\noindent\textbf{Funding.}
This work was supported by JST PRESTO (JPMJPR1905), Japan Society for the Promotion of Science (21K18726, and 21H01848), Cabinet Office, Government of Japan (Subsidy for Reg. Univ. and Reg. Ind. Creation), Research Foundation for Opto-Science and Technology, KDDI Foundation, Telecommunications Advancement Foundation, and "R\&D of high-speed THz communication based on radio and optical direct conversion" (JPJ000254) made with the Ministry of Internal Affairs and Communications of Japan.

\medskip
\noindent\textbf{Disclosures.} The authors declare no conflicts of interest.

\medskip
\noindent\textbf{Supplemental document} See Supplement 1 for supporting content.
\bibliography{sample}

\bibliographyfullrefs{sample}


\ifthenelse{\equal{\journalref}{aop}}{%
\section*{Author Biographies}
\begingroup
\setlength\intextsep{0pt}
\begin{minipage}[t][6.3cm][t]{1.0\textwidth} 
  \begin{wrapfigure}{L}{0.25\textwidth}
    \includegraphics[width=0.25\textwidth]{john_smith.eps}
  \end{wrapfigure}
  \noindent
  {\bfseries John Smith} received his BSc (Mathematics) in 2000 from The University of Maryland. His research interests include lasers and optics.
\end{minipage}
\begin{minipage}{1.0\textwidth}
  \begin{wrapfigure}{L}{0.25\textwidth}
    \includegraphics[width=0.25\textwidth]{alice_smith.eps}
  \end{wrapfigure}
  \noindent
  {\bfseries Alice Smith} also received her BSc (Mathematics) in 2000 from The University of Maryland. Her research interests also include lasers and optics.
\end{minipage}
\endgroup
}{}

\end{document}